\documentclass[12pt]{article}
\usepackage{amssymb,amsmath,amsthm,amsfonts,amscd}
 \usepackage{graphicx}
\newcommand\be{\begin{equation}}
\newcommand\ee{\end{equation}}
\newcommand\bea{\begin{eqnarray}}
\newcommand\eea{\end{eqnarray}}

\newcommand{\fatalpha}{{\bf \alpha \kern -0.44em \alpha}}
\newcommand{\fatsigma}{{\bf \sigma \kern -0.54em \sigma}}
\newcommand{\tpchi}{{\bf D \kern -0.35em D}}
\newcommand{\llambda}{{\bf \lambda \kern -0.45em \lambda}}

\bibliography{plain}
\title{\bf \Large{MAGNETO-ACOUSTIC WAVE OSCILLATIONS IN SOLAR SPICULES}} \vspace{20mm}
\author{}

\pagebreak

\author{A. Ajabshirizadeh$^{1,2,3}$, E. Tavabi$^{1,4}$, S. Koutchmy$^{4}$ \\$^{1}$ Department of Theoretical Physics and Astrophysics, Tabriz University, 51664\\ Tabriz,\\
$^{2}$ Research Institute for Astronomy and Astrophysics of
Maragha,(RIAAM),\\$^{3}$ Research Institute for Applied Physics and
Astronomy of Khadjeh Nassiraldin,\\ Iran,\\$^{4}$ Institut
d'Astrophysique de Paris and UPMC, 98 Bis Boulevard Arago, F-75014\\
Paris, France.\\(E-mail:  a-adjab@tabrizu.ac.ir,
tavabi@tabrizu.ac.ir , koutchmy@iap.fr).}

\vspace{20mm}

\begin{document}
\maketitle \vspace{15mm}
\newpage

%%%%%%%%%%%%%%%%%%%%%%%%%%%%%%%%%%%%%%%%%%%%%%%%%%%%%%%%%%%%%%%%%%%%%%%%%%%%%%%%%%%%%%%%%%%%%%%%%%%%%%%%%%%%%%%%%%%%%%%%%%%
%%%%%%%%%%%%%%%%%%%%%%%%%%%%%%%%%%%%%%%%%%%%%%         ABSTRACT        %%%%%%%%%%%%%%%%%%%%%%%%%%%%%%%%%%%%%%%%%%%%%%%%%%%%
%%%%%%%%%%%%%%%%%%%%%%%%%%%%%%%%%%%%%%%%%%%%%%%%%%%%%%%%%%%%%%%%%%%%%%%%%%%%%%%%%%%%%%%%%%%%%%%%%%%%%%%%%%%%%%%%%%%%%%%%%%%

\begin{abstract}
Some observations suggest that solar spicules show small amplitude
and high frequency oscillations of magneto-acoustic waves, which
arise from photospheric granular forcing. We apply the method of MHD
seismology to determine the period of kink waves. For this purposes,
the oscillations of a magnetic cylinder embedded in a field-free
environment is investigated. Finally, diagnostic diagrams displaying
the oscillatory period in terms of some equilibrium parameters are
provided to allow a comparison between theoretical results and those
coming from observations.



\end{abstract}

\newpage
%%%%%%%%%%%%%%%%%%%%%%%%%%%%%%%%%%%%%%%%%%%%%%%%%%%%%%%%%%%%%%%%%%%%%%%%%%%%%%%%%%%%%%%%%%%%%%%%%%%%%%%%%%%%%%%%%%%%%%%%%%%
%%%%%%%%%%%%%%%%%%%%%%%%%%%%%%%%%%%%%%%%%%%%%%    I.     INTRODUCTION     %%%%%%%%%%%%%%%%%%%%%%%%%%%%%%%%%%%%%%%%%%%%%%%%%
%%%%          The subject or topic.A statement of the problem ,etc.Comments on the way it is to be treated.            %%%%
%%%%%%%%%%%%%%%%%%%%%%%%%%%%%%%%%%%%%%%%%%%%%%%%%%%%%%%%%%%%%%%%%%%%%%%%%%%%%%%%%%%%%%%%%%%%%%%%%%%%%%%%%%%%%%%%%%%%%%%%%%%
\section{Introduction}
%%%%%%%%%%%%%%%%%%%%%%%%%%%%%%%%%%
%%%%%%%%%%%%%%%%%%%%%%%%%%%%%%%%%%
Spicules are jet-like chromospheric structures and are usually seen
all around the limb of the Sun arising in different directions. The
mechanism of spicule formation and evolution is not well understood
(for the propulsive mechanisms, see reviews of Sterling 2000;
Lorrain and Koutchmy 1996; Filippov et al. 2006). Spicules are
relatively homogeneous in height along their life time of
approximately 5-15 min., i.e. they are short-lived and comparable to
the photospheric granules lifetime. They have typical up flow speeds
of 20-50km/s, spicules diameter at chromospheric layers are of the
order of 200-500km. The mean number of spicules per supergranule
cells at height of   km is approximately 40 (Pataraya et al. 1990),
which is covered about 1 percent of the Sun's surface and they are
usually concentrated between supergranule cells. Their temperatures
and density are higher than those of the surrounding
environment,Parenti et al. (1999) estimated $n_{e}=10^{10}cm^{-3}$
and $T_{e}\approx2\times10^{5}$ $^{\circ}K$ in giant spicules see
Koutchmy and Loucif, also named macrospicules, are observed over
20,000 km off-limb and live 40 min. (Xia et al. 2005)  in case of
macro-spicules. This means that a magnetic field of 10G or more is
needed for the low-$\beta$ (ratio of the thermal pressure to the
magnetic pressure) conditions (Wilhelm 2000)  in case of
macro-spicules. Spicule usually have oscillation behavior, the
existence of 5 minutes oscillations in spicules have been firstly
reported by Kulizhanishvili and Nikolsky (1978) and others including
spectroscopically resolved observations. Recently image sequences
were studied by De Pontieu et al. 2003, 2004; Xia et al. (2005);
Ajabshirizadeh et al. (2007). These oscillations seem to be related
to p-modes, but it is evident that if spicules are driven by
p-modes, crucial details about their formation are still missing.
Clearly, not all spicular flows are periodic, whereas most
photospheric oscillations are. In addition, the horizontal scale for
amplitude coherence of p-modes ($\approx 8000 km$) is well beyond
the width of fibrils (De Pontieu et al. 2003). On the-other-hand,
oscillations in spicule with shorter period have been reported by
Nikolsky and Platova (1971). They found that spicules oscillate
along the limb with a characteristic period of about 1 min. If
spicules are formed in thin magnetic flux tubes, then the periodic
displacement of the axis observed by them was probably due to the
propagation of kink waves. More recently, Kukhianidze et al. 2006
have reported the observational signature of propagating kink waves
in spicules. The period of waves was estimated to be 35-70s for a
spicule with height 3500 km which may carry photospheric mechanical
energy into the corona. The cutoff period of kink waves due to
stratification in the hydrostatic photosphere is $\approx660s$ so
the expected period of kink waves is well below the cutoff value
(Singh and Dwivedi, 2007). The wavelength was found to be
$\approx1000km$ at the photosphere level which indicated a granular
origin of the waves. Magnetic flux tubes support transverse kink
waves that can be generated in photospheric magnetic flux tubes
through buffeting action of granular motions (Roberts, 1979; Hasan
and Kalkofen, 1999) although the extrapolation of the network
magnetic field toward the corona is still a matter of discussion. In
this study, the effect of gravitational stratification has been
ignored, Singh and Dwivedi 2007 have considered kink mode periods
with this effect and confirms with our results.

%%%%%%%%%%%%%%%%%%%%%%%%%%%%%%%%%%%%%%%%%%%%%%%%%%%%%%%%%%%%%%%%%%%%%%%%%%%%%%%%%%%%%%%%%%%%%%%%%%%%%%%%%%%%%%%%%%%%%%%%%%
%%%%%%%%%%%%%%%%%%%%%%%%%%%%%%%%%%%%%%%%%%%%%%   II.     DEVELOPMENT      %%%%%%%%%%%%%%%%%%%%%%%%%%%%%%%%%%%%%%%%%%%%%%%%%
%%%%%%%%%%%%%%%%%%%%%%%%%%%%%%%%%%%%%%%%%%%%%%%%%%%%%%%%%%%%%%%%%%%%%%%%%%%%%%%%%%%%%%%%%%%%%%%%%%%%%%%%%%%%%%%%%%%%%%%%%%%

                                         %%%%%%%%%%%%%%%%%%%%%%%%%%%%%%%%%%%%%%%%%%%%%
                                         %   PRESENTATION , ANALYSIS AND DISCUSSION  %
                                         %   (      Involving comments on            %
                                         %      'Advantages' and 'Disadvantages' )   %
                                         %%%%%%%%%%%%%%%%%%%%%%%%%%%%%%%%%%%%%%%%%%%%%
\section{Basic MHD Equations}
\subsection{The Dispersion Relations} Magneto-hydrodynamics (MHD) is one of the key tools to understand the hydrodynamics of magnetized plasmas. It concerns virtually all phenomena observed in the solar atmosphere: coronal loops, filaments, spicules, etc.
Thanks to high spatial resolution, image processing, and time
cadence capabilities of the SoHO and TRACE spacecraft, oscillating
loops (and spicules) and propagating waves have been identified and
localized in the transition region (TR) and chromosphere, and
studied in detail since 1996. Using seeing free observations, they
evidently complement what has been studied for a long time at
ground-based, including spectroscopic analysis. These discoveries
established a new discipline has become know as solar atmosphere
seismology. Many astrophysical plasmas are characterized by a set of
equations that is called ideal MHD equations and includes the
continuity, the momentum, Maxwell's equations and Ohm's law. To
understand the various oscillations and waves we observe in the
spicule plasma we have to find wave solutions of the MHD equations
(Roberts, 1981), the existence of wave solutions is generally
derived by introducing a small perturbation of physical parameters
(such as: density, velocity and magnetic field) of the plasma, and
to derive dispersion relations $\omega(k)$, which tell us either the
group velocity or phase speed of wave. The cylindrical flux tube
appearance of many magnetic structures in the low-$\beta$ plasmas of
the magnetosphere and more specifically the solar chromosphere and
corona encourages an investigation of propagation in cylindrical
geometries. Edwin and Roberts (1983) consider a uniform cylinder of
magnetic field $B_{0}\widehat{z}$ confined to a region of radius 2b,
surrounded by a uniform magnetic field $B_{e}\widehat{z}$, the gas
pressure and density within the cylinder are $P_{0},\rho_{0}$ ,
outside $P_{e},\rho_{e}$ respectively (see figure 1).
 The Fourier form of the velocity disturbance $v_{1}$ in cylindrical coordinate is:
\par

\begin{equation}\label{dens}
\begin{array}{c}
  v_{1}=v_{1}(r)\exp[i(wt+n\theta-k_{z}z)],
\end{array}
     \end{equation}
where n is integer (n=0, 1, 2 …) which describes the azimuthally
behavior of the oscillating tube i.e. the cylindrically symmetric
(sausage or pulsation mode given by n=0, the asymmetric (kink or
taut-wire) mode given by n=1 and higher mode with n=2, 3… are called
the fluting or interchange modes. The general governing equation is
(Edwin and Roberts 1983):
\begin{equation}\label{dens}
\begin{array}{c}
\frac{d}{dr}[\frac{\rho_{\alpha}(r)(k^{2}_{z}V^{2}_{A\alpha}-\omega^{2})}{(m^{2}_{\alpha}+\frac{n^{2}}{r^{2}})}\frac{1}{r}\frac{d}{dr}(rv_{1})]-{\rho_{\alpha}(r)(k^{2}_{z}V^{2}_{A\alpha}-\omega^{2})}v_{1}=0,
\end{array}
     \end{equation}
where $m_{0}$  and $m_{e}$  are defined by $m_{\alpha}$ (with
$\alpha=0$  or $\alpha=e$  inside or outside of the tube
respectively):
\begin{equation}\label{dens}
\begin{array}{c}
m_{\alpha}=\frac{(k^{2}_{z}C^{2}_{\alpha}-\omega^{2})(k^{2}_{z}V^{2}_{A\alpha}-\omega^{2})}{(C^{2}_{\alpha}+V^{2}_{A\alpha})(k^{2}_{z}C^{2}_{T\alpha}-\omega^{2})}
\end{array}
     \end{equation}
where $C_{\alpha}=(\frac{\gamma
P_{\alpha}}{\rho_{\alpha}})^{\frac{1}{2}}$ and
$V_{A\alpha}=\frac{B_{\alpha}}{(\mu\rho_{\alpha})^{\frac{1}{2}}}$
are the sound and Alfvén speed inside (or outside) the cylinder, and
$C_{T\alpha}$ is defined as:
\begin{equation}\label{dens}
\begin{array}{c}
C_{T\alpha}=\frac{C_{\alpha}V_{A\alpha}}{\sqrt{(C^{2}_{\alpha}+v^{2}_{A\alpha})}}
\end{array}
     \end{equation}
( $\gamma$ is the ratio of specific heats.) The external and
internal solutions of MHD equations need to be matched at the
boundary by the continuity of pressure and the perpendicular
component of velocity. After some algebra one gets the dispersion
relation for magneto-acoustic waves in a cylindrical magnetic flux
tube is found to be (Edwin and Roberts 1983; Díaz et al. 2002):
\begin{equation}\label{dens}
\begin{array}{c}
\rho_{e}(\omega^{2}-k^{2}_{z}v^{2}_{Ae})m_{0}\frac{I'_{n}(m_{0}b)}{I_{n}(m_{0}b)}+\rho_{0}(\omega^{2}-k^{2}_{z}v^{2}_{A0})m_{e}\frac{K'_{n}(m_{e}b)}{K_{n}(m_{e}b)}=0,
\end{array}
     \end{equation}
where $I_{n}$ and $K_{n}$ are modified Bessel functions of order n,
with $I'_{n}$ and $K'_{n}$ being the derivatives with respect to the
argument x. This dispersion relation describes both surface (for
$m^{2}_{0}>0$ ) and body waves (for $m^{2}_{0}<0$ ).

\subsection{Kink-mode period}
Magneto-acoustic oscillations of kink mode have now been directly
observed in $H\alpha$  line using 53-cm coronagraph the Abastumani
Astrophysical Observatory at different heights above the photosphere
(Kukhianidze et al. 2006). The ratio of the spicule width 2b to the
spicule full length 2L is $\frac{b}{L}=0.01-0.4$  , which is
correspond to the dimensionless wave number (kL).\\ Therefore, the
observed kink-mode oscillations are in the long-wavelength regime of
$KL<<1$, where the phase speed of kink-mode is practically equal to
the kink speed (Spruit, 1981; Roberts et al. 1984) given by:
\begin{equation}\label{dens}
\begin{array}{c}
C_{k}=(\frac{\rho_{0}V^{2}_{A0}+\rho_{e}V^{2}_{Ae}}{\rho_{0}+\rho\rho_{e}})^{\frac{1}{2}},
\end{array}
     \end{equation}
In the low-$\beta$  plasma limit and for the field free environment,
the expression for the kink speed $C_{k}$ reduces to:
\begin{equation}\label{dens}
\begin{array}{c}
C_{k}\approx(\frac{2}{1+\frac{\rho_{e}}{\rho_{0}}})^{\frac{1}{2}}V_{A0},
\end{array}
     \end{equation}
If we denote the full spicule length l=2L, the wavelength of the
fundamental standing wave is the double spicule length (due to the
forward and backward propagation) i.e., $\lambda=2l$   and thus the
wave number of the fundamental mode (N=1) is
$k_{z}=\frac{2\pi}{\lambda}=\frac{\pi}{l}$, while higher harmonics
(N=2, …) would have wave numbers $k_{z}=N(\frac{\pi}{l})$, then the
time period P of a kink-mode oscillation at the fundamental harmonic
is:
\begin{equation}\label{dens}
\begin{array}{c}
P\approx\frac{2l}{C_{k}}=\frac{2l}{V_{A0}}(\frac{1+\frac{\rho_{e}}{\rho_{0}}}{2})^{\frac{1}{2}},
\end{array}
     \end{equation}
and for higher harmonics,
\begin{equation}\label{dens}
\begin{array}{c}
P\approx\frac{2l\sqrt{\mu}}{NB_{0}}(\frac{\rho_{0}+\rho_{e}}{2})^{\frac{1}{2}}.
\end{array}
     \end{equation}

%%%%%%%%%%%%%%%%%%%%%%%%%%%%%%%%%%%%%%%%%%%%%%%%%%%%%%%%%%%%%%%%%%%%%%%%%%%%%%%%%%%%%%%%%%%%%%%%%%%%%%%%%%%%%%%%%%%%
%%%%%%%%%%%%%%%%%%%%%%%%%%%%%%%%%%%%%%%%%%%   III.   CONCLUSION       %%%%%%%%%%%%%%%%%%%%%%%%%%%%%%%%%%%%%%%%%%%%%%
%%%%%%%%              A summary of the main points in II. Self views/opinions and decisions               %%%%%%%%%%
%%%%%%%%%%%%%%%%%%%%%%%%%%%%%%%%%%%%%%%%%%%%%%%%%%%%%%%%%%%%%%%%%%%%%%%%%%%%%%%%%%%%%%%%%%%%%%%%%%%%%%%%%%%%%%%%%%%%

\section{Results}
Such as has been pointed out before, the determinant coming from
dispersion equation (2-5) must be truncated by taking into account a
finite number of basis functions and we will use surface wave
($m^{2}_{0}>0$ in Eq. 2-5). As we know, the kink wave is essentially
non-dispersive and has a phase speed equal to the kink speed,
$v_{ph}=\frac{\omega}{k_{z}}\approx C_{k}\approx V_{A0}$ so we
introduce dimensionless frequency which is given by $\frac{\omega
L}{V_{A0}}$ for odd modes and the fundamental modes and their
harmonics have a
cutoff frequency for odd modes equal $\pi$ (see Diaz et al. 2002).\\
In figure 2 the eigenfrequencies of the kink (n=1) odd modes
($\omega_{cutoff}=\pi$ ) have been plotted in term of the spicule
half-thickness. In this plot we can see that for small value of
$\frac{b}{L}$ only the fundamental and lower harmonics could appear
and for large ratio of $\frac{b}{L}$ the frequency of the
fundamental mode is insensitive to the spicule thickness. One of the
essential results of these plots is that the oscillatory frequency
is quite insensitive to the exact value of the ratio $\frac{b}{L}$
(the ratio of the spicule thickness to the half-length of magnetic
field lines inside the spicule), i.e., for a given length 2L,
spicules with different thickness oscillate with the same frequency,
this is perhaps an interesting result, and this subject have been
firstly reported by Díaz et al. 2002, for kink waves in the
prominence fibrils. We continue by choosing two typical values of
the spicule full length, namely, l=3500 and 8500km, and then plot
the frequency of odd modes below the cutoff as a function of
$(\frac{\rho_{e}}{\rho_{0}})$ (see figure 3-a, b).\\
In these plots, we keep $\frac{b}{L}=0.05,0.03$ constant for two
lengths of spicule respectively. For all harmonic modes the
frequency is seen to slowly decrease with increasing
$(\frac{\rho_{e}}{\rho_{0}})$ and more oscillatory modes are present
for large values of this quantity. To compute the corresponding
periods for each harmonic modes, we used l=3500 and 8500km,
$B_{0}=30G$ and $\rho_{0}=3\times10^{-10}kgm^{-3}$. The periods
obtained from these quantities and using Eq. (2-9) is labeled P and
correspond to
the right vertical axes in figure 3 (see also Table 1).\\
 The period, P, (figure 3- a, b right vertical axes and table 1) for
different values of $\frac{b}{L}$ , L calculated from the expression
of Eq. (2-5) and (2-9), show that the fundamental and first harmonic
have typical periods of $\approx 30-80s$. These periods which are
obtained from a simple MHD approach are found in the more recently
reported observational results (Kukhianidze et al. 2006). where
oscillation periods in the range of kink wave of $\approx 35-70s$
are found for spicules with height 3500-8700 km. From the
theoretical point of view, we therefore expect kink wave periods in
the range of $\approx 80-120s$ for length of $\approx
10,000-14,000km$ which could be found from future observations of
kink
wave inside spicules (Hinode observations).\\
 \emph{\textbf{ Acknowledgments.}} The authors are most grateful to A. J. Díaz for useful discussions and
critical reading of the manuscript.

%%%%%%%%%%%%%%%%%%%%%%%%%%%%%%%%%%%%%%%%%%%%%%%%%%%%%%%%%%%%%%%%%%%%%%%%%%%%%%%%%%%%%%%%%%%%%%%%%%%%%%%%
%%%%%%%%%%%%%%%%%%%%%%%%%%%%%%%%%%%%%%%%%%%%%%%%%%%%%%%%%%%%%%%%%%%%%%%%%%%%%%%%%%%%%%%%%%%%%%%%%%%%%%%%

\newpage
{\bf Figure Captions}

 {\bf Figure-1:} Fig.1- Sketch of the equilibrium configuration used in this study. The density and magnetic field inside the spicule are $B_{e},\rho_{e}$,
  and in chromosphere environment are $B_{0},\rho_{0}$. The magnetic field is uniform and along z-axis..

{\bf Figure-2:} Fig. 2. Variation of the frequency with the spicule
half-thickness for Kink-Modes for the set of value 2L=8500km, and
for the density ratio $\frac{\rho_{e}}{\rho_{0}}$=0.03.

 {\bf Figure-3:} Fig. 3- a, b. Frequency of the kink odd modes vs.$\frac{\rho_{e}}{\rho_{0}}$ for
two full length of sipules and $\frac{b}{L}=0.05,0.03$ for the
parameters . The right-hand axis provides the period P after
estimation that magnetic field strength, the spicule density, and
full-length of spicules are $\rho_{0}=3\times10^{-10}kgm^{-3}$
(corresponding to a $2\times10^{11}$ number density), $B_{0}=30G$,
and 2L=3500, 8500km, respectively.

{\bf Figure-4:} TABLE 1

%%%%%%%%%%%%%%%%%%%%%%%%%%%%%%%%%%%%%%%%%%%%%%%%%%%%%%%%%%%%%%%%%%%%%%%%%%%%%%%%%%%%%%%%%%%%%%%%%%%%%%%%%%%%%%%%%%%%%%%%%
%%%%%%%%%%%%%%%%%%%%%%%%%%%%%%%%%%%%%%%%%%%%%%%%%        END        %%%%%%%%%%%%%%%%%%%%%%%%%%%%%%%%%%%%%%%%%%%%%%%%%%%%%
\end{document}